\documentclass[twocolumn,prb,showpacs]{revtex4}

\usepackage{epsfig}
\usepackage{graphics}
\usepackage{graphicx}
\usepackage{dcolumn}
\usepackage{bm}

\def\scbo{SrCu$_{2}$(BO$_{3}$)$_{2}$\,}
\def\cm-1{cm$^{-1}$}

\begin{document}

\title
{
Symmetry and light coupling to phononic and collective magnetic 
excitations in \scbo
}

\author{A.~Gozar$^{1,2,*}$}
\author{B.S. Dennis$^{1}$}
\author{H. Kageyama$^{3}$}
\author{G.~Blumberg$^{1,\dag}$}

\affiliation{
$^{1}$Bell Laboratories, Lucent Technologies, Murray Hill, NJ 07974 \\
$^{2}$University of Illinois at Urbana-Champaign, Urbana, IL 61801-3080\\
$^{3}$Department of Chemistry, Graduate School of Science, Kyoto 
University, Kyoto 606-8502 Japan}

\date{\today}

\begin{abstract}
We perform a low temperature Raman scattering study of phononic and 
collective spin excitations in the orthogonal dimers compound \scbo, 
focussing on the symmetry and the effects of external fields on the 
magnetic modes.
The zero field symmetry and the behavior in magnetic fields of the 
elementary and bound magnetic triplet states are experimentally 
determined.
We find that a minimal 4-spin cluster forming the unit cell is able 
to describe the symmetry as well as the anisotropic dispersions in 
external fields of the spin gap multiplet branches around 24~\cm-1.
We identify two Raman coupling mechanisms responsible for the 
distinct resonance behavior of these magnetic modes and we show that 
one of these can be ascribed to an effective intra-dimer 
Dzyaloshinskii-Moriya spin interaction.
Our data also suggest a possible explanation for the existence of a 
strongly bound two-triplet state in the singlet sector which has an 
energy below the spin gap.
The low temperature phononic spectra suggest strong spin-phonon 
coupling and show intriguing quasi-degeneracy of modes in the context 
of the present crystal structure determination.
\end{abstract}


\maketitle

\section{Introduction}
Several properties make \scbo a unique system among the known quantum magnets.
This compound is a realization of a two-dimensional (2D) spin system 
with a disordered ground state even at very low temperatures and a 
spin gap $\Delta \approx$~24~\cm-1 (3~meV)  from the singlet ($S = 
0$) ground state to the lowest excited triplet ($S = 1$) 
state~\cite{MiyaharaJPCM03,KageyamaPRL99}.
The strengths of the relevant magnetic interactions place this 
compound close to a quantum critical point (QCP).
Moreover, data in high magnetic fields show plateaus at commensurate 
(1/8, 1/4 and 1/3) values of the saturation 
magnetization~\cite{KageyamaPRL99,OnizukaJPSJ00,KodamaScience02}.
The plateau states can be thought of as crystalline arrangements of 
magnetic moments separating regions of continuous rise in 
magnetization, the latter allowing for an interpretation in terms of 
Bose-Einstein condensation of triplet 
excitations~\cite{RiceScience02}.
It has also been suggested~\cite{SriramPTP02} that doping in this 
system (regarded as a Mott-Hubbard insulator) may lead to a 
superconducting phase mediated by antiferromagnetic (AF) 
fluctuations, a mechanism similar in spirit to one of the scenarios 
proposed for the high T$_{c}$ cuprates~\cite{AndersonScience87}.

The $S = 1/2$ Cu spins are arranged in weakly coupled 2D layers 
defining the $(ab)$ plane.
In each of these sheets they form orthogonal spin-dimer 
lattices~\cite{SmithJSSC91,SpartaEPJB01}, see Fig.~1a.
At T$_{c}$~=~395~K the system undergoes a second order phase 
transition from the space group $I4/mcm$ to $I\bar{4}2m$ on cooling 
down from the high temperature side.
In the $I4/mcm$ phase the planes containing the Cu atoms are flat and 
they form mirror symmetry elements.
The transition at 395~K can be understood as the buckling of the Cu 
planes which lose their mirror symmetry property.
Inversion symmetry is lost as well below T$_{c}$ but the number of 
atoms in the unit cell remains unchanged due to the orthogonality of 
the spin-dimer network.
Because of the same orthogonal arrangement, there are two spin dimers 
in the unit cell.
As a result, the spin gap excitation, defined as the transition from 
the $S = 0$ ground state to that (usually the lowest in energy) $S = 
1$ level whose wavefunction contains a single spin dimer excited to 
the triplet state \cite{MiyaharaJPCM03,KnetterPRL00,KnetterPRL04}, 
has a 'fine structure' made out of six levels, three from each dimer 
in the unit cell.
\begin{figure}[b]
\centerline{
\epsfig{figure=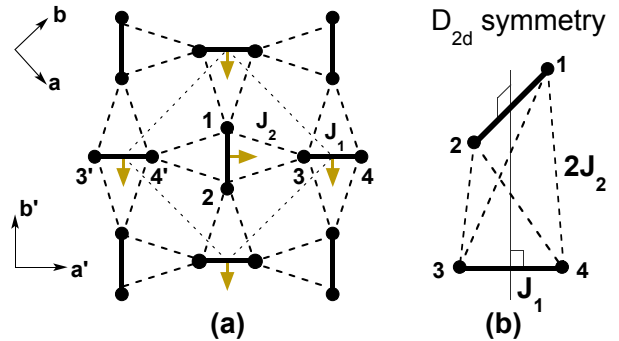,width=80mm}
}
\caption{
(Color online)
(a) The structure of the spin-dimer compound \scbo.
The unit cell is shown by the short dashed square.
The circles are the $S = 1/2$ Cu spins, the solid and long dashed 
lines represent the intra-dimer ($J_{1}$) and inter-dimer ($J_{2}$) 
AF superexchange interactions.
The arrows correspond to the proposed intra-dimer antisymmetric spin 
interaction $d_{ab}$.
Below 395~K the orthogonal dimers are no longer coplanar.
(b) The 4-spin cluster corresponding to the unit cell in (a).
As expected, the symmetry of this object is $D_{2d}$, which is the 
point group associated with the crystal space group of \scbo below 
395~K.
Note that the inter-dimer superexchange $J_{2}$ is effectively 
doubled if periodic boundary conditions are used.}
\label{f1}
\end{figure}

The 2D spin-dimer lattice can be described well by taking into 
account the nearest and next nearest neighbor AF super-exchange 
interactions $J_{1}$ and $J_{2}$ respectively, see
Fig.~1a.
In this approximation and for a ratio $x = J_{2}/J_{1}$ lower than 
about 0.7, the direct product of singlet dimers on each rung, which 
is always an exact eigenstate of the Hamiltonian, is also the ground 
state of the system and the spin gap is finite although it gets 
renormalized down with increasing $x$ due to many body 
effects~\cite{KnetterPRL00,MiyaharaJPSJS00}.
For large $x$ the ground state changes and the Hamiltonian has long 
range AF order, other possible intervening states separated by QCP's 
being proposed to exist around 0.7, see~Ref.~\cite{MiyaharaJPCM03}
Based on fits to magnetization data, predictions of symmetries, 
relative energies and dispersions of single and composite triplet 
excitations, theoretical estimates for $x$ range from 
0.603~Refs.\cite{KnetterPRL00,KnetterPRL04} to about 
0.68~Refs.\cite{MiyaharaPRL99,MiyaharaJPSJS00}

An analysis of the vibrational modes is of interest in \scbo because 
spin-lattice interaction has been suggested to be relevant to the 
magnetic dynamics at low temperatures and high magnetic fields.
In particular, spin-phonon interaction has been invoked in order to 
explain the selection rules of the magnetic transitions seen in 
infra-red (IR) absorption~\cite{ToomasPrivate}.
The coupling between the lattice and magnetic degrees of freedom was 
also taken into account in order to describe the spin density profile 
at high fields in the magnetization plateau 
states~\cite{KodamaScience02}.
So far the study of phononic excitations has been related mostly to 
the crystallographic changes at 395~K.
It has been established in Refs.~\cite{SpartaEPJB01,ChoiPRB03} that 
the structural transition displays soft mode behavior (see 
Refs.~\cite{SpartaEPJB01,ChoiPRB03} also for a phononic symmetry 
analysis).
The phononic mode which condenses at 395~K belongs to the B$_{1u}$ 
representation of the high temperature symmetry group and, in terms 
of Cu atoms, it involves essentially an alternate displacement along 
the $c$-axis of the nearest neighbor dimers.
Interestingly, several phonons appearing below T$_{c}$ were observed 
in $(ca)$ polarization as very close in energy shoulders of some of 
the modes found above the transition.
Due to the absence of inversion symmetry below 395~K this 
experimental fact suggested the existence of  almost degenerate even 
and odd (transforming like the E$_{g}$ and E$_{u}$ representations 
respectively) excitations in the high temperature 
phase~\cite{ChoiPRB03}.

In terms of magnetic properties, elementary and two-triplet 
excitations have been studied by inelastic neutron scattering 
(INS)~\cite{KageyamaPRL00,CepasPRL01}, electron spin resonance 
(ESR)~\cite{NojiriJPSJ99,NojiriJPSJ03} and IR absorption 
spectroscopy~\cite{ToomasPrivate,ToomasPRB00}.
The low temperature Raman scattering results shown in 
Ref.~\cite{LemmensPRL00} are focussed on composite magnetic 
excitations in the $S = 0$ channel and are used to extract 
quantitatively the ratio $x = J_{2} / J_{1}$.
In our Raman study, the emphasis is however on the elementary and 
composite collective excitations in the $S = 1$ sector, the main 
results being related to the experimental determination of their 
symmetries and anisotropic behavior in magnetic fields.
We are also able to relate some of these new findings to results of a 
4-spin cluster analysis as well as to propose an additional effective 
spin interaction induced by spin-orbit coupling.
It has been established that due to the frustrated nature of the 
magnetic interactions, the one triplet excitations are local, weakly 
dispersive in the reciprocal space, while two-particle states are 
more mobile~\cite{KageyamaPRL00} and have contributions from the 
whole Brillouin zone~\cite{KnetterPRL00,KnetterPRL04}.

In spite of a lot of experimental and theoretical effort for 
understanding the magnetic properties there are several open 
questions.
An exact determination of the ratio of the exchange interactions $x$, 
which is important due to the proximity to the QCP, is still missing.
One interesting aspect in this regard is the observation of a 
magnetic state at 21.5~\cm-1, which is \emph{below} the energy of the 
spin gap multiplet $\Delta$.
This could be seen in ESR~\cite{NojiriJPSJ03}, INS~\cite{CepasPRL01} 
and IR~\cite{ToomasPrivate} data in energy level anti-crossings in 
the downward dispersions with magnetic field of some of the spin gap 
branches.
The existence of such a low energy mode brings into question the 
exact quantitative estimation of the AF exchange parameters.
As we will show, a direct comparison between our experimental 
findings and theoretical predictions regarding the symmetry of these 
modes is illuminating in this respect.

A different set of questions is related to the way the external 
radiation field couples to the magnetic excitations.
While the photon induced spin-exchange process insures the Raman 
coupling to $S = 0$ two-triplet 
excitations~\cite{LemmensPRL00,FleuryPR68}, the transitions to $S = 
1$ states require the presence of spin-orbit 
coupling~\cite{FleuryPR68}.
Although an effective antisymmetric Dzyaloshinskii-Moriya (DM) term 
originating in the spin-orbit interaction (the $H^{DM}_{c}$ 
contribution explicitly written in Eq.~\ref{eq3}) has been proposed 
to explain the 'fine structure' of the six levels forming the gap
multiplet around 24~\cm-1 seen in a neutron scattering study, 
Ref.~\cite{CepasPRL01}, this term does not mix the singlet ground 
state to excited $S = 1$ states.
As a result, the nature of the coupling leading to the observed ESR 
and IR data is still to be understood.
Possible candidates were suggested in the discussion of ESR 
data~\cite{NojiriJPSJ03} and dynamical spin-phonon induced effective 
DM interactions have been invoked in order to explain the optical
absorption spectra~\cite{ToomasPrivate,ToomasPRB00}.

In this article we study phononic modes and collective magnetic 
excitations in \scbo with the experimental emphasis placed on the 
understanding of the magnetic Raman scattering in the $S~=~1$ channel.
In the low temperature in-plane polarized phononic spectra we find 
(in the 50~-~350~\cm-1 energy range) several pairs of modes with 
similar energies.
This quasi-degeneracy is intriguing because group theory analysis 
suggests that they are related to different atomic vibrational 
patterns.
Regarding the magnetic Raman scattering data, our novel findings are 
the following.
The symmetry analysis and exact diagonalization of the 4-spin system 
Hamiltonian (see Fig.~1b) given by:
\begin{equation}
H = H_0 + H^{DM}_{c} + H^{DM}_{ab}
\label{eq1}
\end{equation}
which includes the main, unperturbed, Heisenberg term, $H_{0}$, the 
antisymmetric inter-dimer DM term proposed in Ref.~\cite{CepasPRL01}, 
$H^{DM}_{c}$, and an additional intra-dimer DM term, $H^{DM}_{ab}$, 
reading:
\begin{eqnarray}
H_0= J_{1} \sum_{(i,j) \ NN} {\bf S}_{i} \cdot {\bf S}_{j} + J_{2} 
\sum_{(i,j) \ NNN} {\bf S}_{i} \cdot {\bf S}_{j},
\label{eq2} \\
H^{DM}_{c} =\sum_{(i,j) \ NNN} \vec{d}_{c}^{\ (i,j)}  ({\bf S}_{i} 
\times {\bf S}_{j}),
\label{eq3} \\
H^{DM}_{ab} =\sum_{(i,j) \ NN} \vec{d}_{ab}^{\ (i,j)}  ({\bf S}_{i} 
\times {\bf S}_{j})
\label{eq4}
\end{eqnarray}
explains the experimentally determined symmetries of the zero field 
Brillouin zone center spin gap branches around 24~\cm-1 (confirming 
the local nature of the elementary one-triplet modes) but fails to 
account for the two-triplet excitations.
In the equations above $i$ and $j$ are nearest neighbor (NN) or next 
nearest neighbor (NNN) Cu sites and  $\vec{d}_{c}$ and 
$\vec{d}_{ab}$ are inter and intra-dimer antisymmetric spin 
interaction vectors.
Remaining confined to the 4-spin cluster we find that by introducing 
the effective intra-dimer DM interaction $H^{DM}_{ab}$ we are also 
able to reproduce the observed selection rules and intensity 
variations of the spin gap branches in external magnetic fields.
These selection rules also require that the energy of the $S = 0$ 
two-triplet bound state formed from spins confined within a unit cell 
is \emph{below} $\Delta$ (in the 4-spin cluster this is equivalent to 
$x \geq 0.5$ in Fig.~4) suggesting a high binding energy for this two 
particle excitation.
In the last section, we also show results of a resonance study which 
allows us to identify the action of two different light coupling 
mechanisms to magnetic excitations.

\section{Experimental}
Raman scattering from the $ab$ surface of a single crystal of \scbo, 
grown as described in Ref.~\cite{KageyamaJCG99}, was performed using 
an incident power density less than 1~mW focussed to a 100~$\mu m$ 
diameter spot.
The crystallographic axes orientation was determined by X-ray diffraction.
The data in magnetic fields, taken at a sample temperature of about 
3~K, was acquired with a continuous flow cryostat inserted in the 
horizontal bore of a superconducting magnet.
We used the $\omega_{L} = $~1.92 and 2.6~eV excitation energies of a 
Kr$^{+}$ laser and a triple-grating spectrometer for the analysis of 
the scattered light.
The data were corrected for the spectral response of the spectrometer 
and CCD detector.

Polarized Raman scattering can probe Brillouin zone center 
excitations that belong to different symmetry representations within 
the space group of the crystal structure.
We denote by $(\textbf{e}_{in} \textbf{e}_{out})$ a configuration in 
which the incoming/outgoing photons are polarized along the 
$\textbf{e}_{in}$/$\textbf{e}_{out}$ directions (see Fig.~1 for axes 
notations).
The $(RR)$ and $(RL)$ notations refer to circular polarizations, 
$\textbf{e}_{in} = (\hat{a} - i \hat{b}) / \sqrt{2}$, with 
$\textbf{e}_{out} = \textbf{e}_{in}$ for $(RR)$ and $\textbf{e}_{out}
= \textbf{e}_{in}^{*}$ for the $(RL)$ geometry.
The point group associated to the high temperature phase space group 
of the \scbo crystal, $I4/mcm$, is $D_{4h}$.
The point group associated to the $I\bar{4}2m$ space group 
corresponding to the lower temperature phase is $D_{2d}$.
In $D_{2d}$, the $(RR)$, $(RL)$, $(aa)$, $(ab)$, $(a'a')$ and 
$(a'b')$ polarizations probe excitations which belong to the A$_{1}$ 
+ A$_{2}$, B$_{1}$ + B$_{2}$, A$_{1}$ + B$_{1}$, A$_{2}$ + B$_{2}$, 
A$_{1}$ + B$_{2}$ and A$_{2}$ + B$_{1}$ irreducible representations 
of $D_{2d}$.

\section{Phononic symmetries}
In Fig.~2 we show six low temperature Raman spectra excited with the 
$\omega_{L} = 1.92$~eV laser  energy.
The arrows point to the observed modes above 60~\cm-1 and also shown 
are their energies and symmetries.
Magnetic fields do not affect their energies which indicates the 
phononic nature of these excitations.
The modes below 60~\cm-1 are not indexed since they will be discussed 
in the next sections.
The excitation at 59~\cm-1 corresponds to the fully symmetric soft 
mode of the structural transition at 
395~K~\cite{SpartaEPJB01,ChoiPRB03}.
The corresponding two-phonon excitation in the A$_{1}$ channel is 
seen at 121.8~\cm-1 and very close to it another sharp mode with 
B$_{1}$ symmetry.
Similarly to the data in $(ca)$ polarization in 
Ref.~\cite{ChoiPRB03}, where below 395~K several new modes appear as 
shoulders of phonons existing above the transition, we observe 
several pairs of modes having close energies.
For example doublet structures are observed around 155~\cm-1, where 
we see a pair of B$_{2}$ and A$_{2}$ excitations, around 285~\cm-1, 
where we observe two modes having B$_{1}$ and A$_{1}$ symmetries, and 
also around 320~\cm-1 where we see a pair made of B$_{2}$ and B$_{1}$ 
symmetric excitations.
\begin{figure}[b]
\centerline{
\epsfig{figure=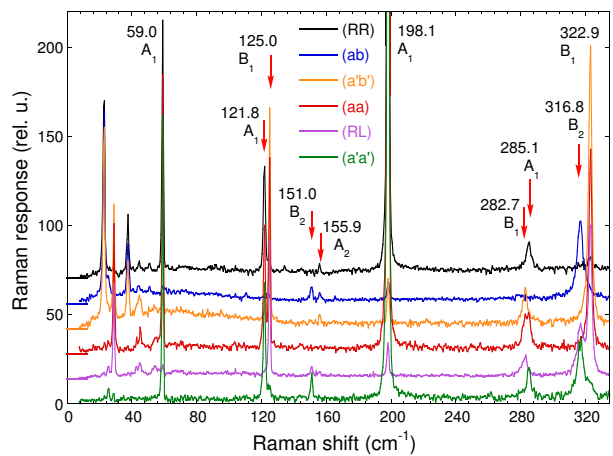,width=80mm}
}
\caption{
(Color online)
Raman spectra at T~=~3~K in six polarization configurations and zero 
applied field.
Next to each vertical arrow we show the energies (in \cm-1) and 
symmetries of the modes above $\approx 60$~\cm-1.
The spectra are vertically offset.
The excitation energy used is $\omega_{L} = 1.92$~eV.
}
\label{f2}
\end{figure}

One way to explain this behavior, suggested in Ref.~\cite{ChoiPRB03}, 
is to assume that in the high temperature phase there are phonons 
which are odd (ungerade: 'u') and even (gerade: 'g') with respect 
to inversion but very close in energy and to try to identify them by 
looking at similar atomic vibrations corresponding to 'u' and 'g' 
representations respectively.
Following this idea in more detail and taking into account that the 
set of irreducible representations 
\{A$_{1g}$,A$_{2g}$,B$_{1g}$,B$_{2g}$,A$_{1u}$,A$_{2u}$,B$_{1u}$,B$_{2u}$\} 
of the $D_{4h}$ point group becomes the set of 
\{A$_{1}$,A$_{2}$,B$_{1}$,B$_{2}$,B$_{1}$,B$_{2}$,A$_{1}$,A$_{2}$\} 
representations (in this order) in $D_{2d}$, we would have for 
example that the (A$_{1}$,B$_{1}$) pair around 284~\cm-1 corresponds 
either to a (A$_{1g}$,A$_{1u}$) group or to a (B$_{1u}$,B$_{1g}$) 
group in the high temperature phase.
This is because we chose gerade-ungerade pairs and, as shown above, 
the A$_{1}$ symmetric mode can originate either from a A$_{1g}$ or a 
B$_{1u}$ symmetric phonon while the B$_{1}$ mode could be either a 
B$_{1g}$ or a A$_{1u}$ phonon in the high temperature phase.
Similar reasoning would suggest that the (B$_{1}$,B$_{2}$) group 
around 320~\cm-1 originates either from a pair of (B$_{1g}$,A$_{2u}$) 
or (B$_{2g}$,A$_{1u}$) above 395~K and also that the origin of the 
(B$_{2}$,A$_{2}$) group around 155~\cm-1 can be pairs of 
(B$_{2g}$,B$_{2u}$) or (A$_{2g}$,B$_{1u}$) modes in the $I4/mcm$ 
phase which has $D_{4h}$ as the associated point group.

We performed a symmetry analysis of the $k = 0$ atomic vibrations in 
the high temperature phase and our conclusion is that the approach 
suggested in Ref.~\cite{ChoiPRB03} does not provide an a priori 
reason for the quasi-degeneracy.
This conclusion, as explained in the following, is based just on a 
simple inspection of the character table of the $D_{4h}$ point 
group~\cite{KosterBook}.
One can easily note that in $D_{4h}$ the even modes are symmetric 
with respect to the mirror symmetry in the Cu(BO$_{3}$) planes while 
the odd vibrations are antisymmetric with respect to this symmetry 
operation.
This means that the 'u' phonons in the high temperature phase 
correspond to vibrations of the atoms along the $c$-axis while the 
'g' modes consist of in-plane movements.
Due to this difference in the oscillation patterns, one parallel and 
one perpendicular to the CuBO$_{3}$ planes, it is hard to explain the 
closeness of phononic energies at this qualitative level.
This is why we find intriguing the observed quasi-degeneracy of the 
phononic modes in the context of the present crystal structure 
determination.

We remark that on the other hand one could find vibrations which 
involve similar oscillations at the 'molecular' level (for instance 
groups of atoms forming the Cu-O spin dimer structure or groups of O2 
atoms bridging nearest neighbor spin dimers) and which belong to 
different group representations because of the different 
'inter-molecular' phase pattern.
We suggest that, remaining within the conclusions of X-ray 
studies~\cite{SpartaEPJB01} which so far have not found evidences for 
additional crystallographic changes at low temperatures (that in turn 
may produce phonon splittings), good candidates for understanding 
this behavior are provided by the inter-dimer BO$_{3}$ molecular 
complexes whose rotations as a whole around the $a'$ and $c$-axes or 
whose in and out of the plane translations may turn out to be similar 
in energies.
The data suggest that theoretical work in this respect could be interesting.

We note that the appearance of the weak 155.9~\cm-1 mode in the 
$(RR)$, $(ab)$ and $(a'b')$ polarizations singles this mode out from 
the other excitations because it belongs to the A$_{2}$ symmetry 
channel. The Raman coupling to this excitation is unusual compared to 
the other modes in the sense that it cannot take place $via$ two 
electric dipole transitions.
We find several magnetic resonances (see Fig.~3) with A$_{2}$ 
symmetry, but the absence of magnetic field effects suggests that 
this mode has a preponderant phononic character.
A finite intensity of such excitation provides direct evidence for 
spin-phonon coupling in \scbo.

\section{Magnetic symmetries}
The irreducible representations of the $D_{2d}$ point group probed by 
six scattering geometries, discussed also in the experimental 
section, are shown in the legend of Fig.~3.
In this figure, the six low temperature Raman spectra from Fig.~2 are 
shown for the frequency region below 60~\cm-1.
Three strong features around 23, 29 and 38 \cm-1, denoted by 
T$_{1b}$, S$_{1}$ and T$_{2}$, are observed and they transform like 
the A$_{2}$, B$_{1}$ and A$_{2}$ representations respectively.
Besides these three modes, we observe several other weaker excitations.
Among them we see a set of three A$_{2}$ symmetric modes denoted by 
T$_{3}$, T$_{4}$ and T$_{5}$.
We also note the presence of the excitations denoted by T$_{1e}$ and 
T$_{1f}$ which have B$_{1}$ and B$_{2}$ symmetries, giving rise to a 
small feature seen around 25.6~\cm-1 in all polarizations except for 
the $(RR)$ configuration.
Table~\ref{modes} contains a summary of the observed excitations 
below the 60~\cm-1 frequency range.
\begin{figure}[t]
\centerline{
\epsfig{figure=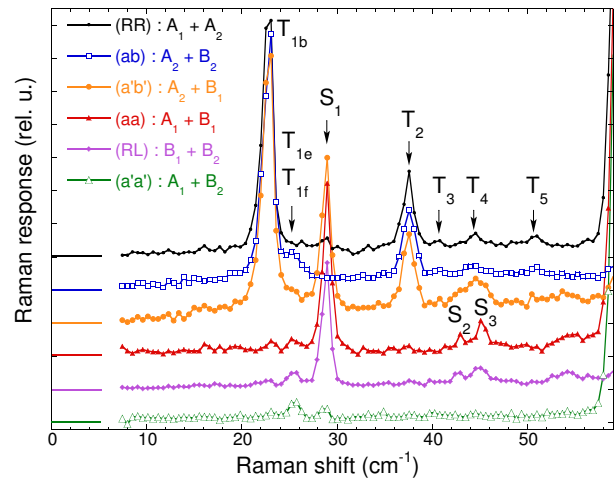,width=80mm}
}
\caption{
(Color online)
Zero field Raman data in taken with $\omega_{L}$~=~1.92~eV 
excitation energy at T~=~3~K in six polarization configurations.
The legend shows the tetragonal symmetries probed in each scattering geometry.
See the text for notations.
}
\label{f3}
\end{figure}

The energies of these Raman active excitations are in agreement with 
those where INS~\cite{KageyamaPRL00}, ESR~\cite{NojiriJPSJ99, 
NojiriJPSJ03} and IR~\cite{ToomasPrivate,ToomasPRB00} data observed 
magnetic modes.
The data in magnetic fields (see the discussion in the following section)
confirm the magnetic nature and the predominant $S = 1$ character of the 'T' modes.
Therefore T$_{1b}$, T$_{1e}$ and T$_{1f}$ excitations modes seem to 
belong to the spin gap multiplet while T$_{2}$, T$_{3}$, T$_{4}$ and 
T$_{5}$ would correspond to multi-particle triplet channels.
\begin{figure}[t]
\centerline{
\epsfig{figure=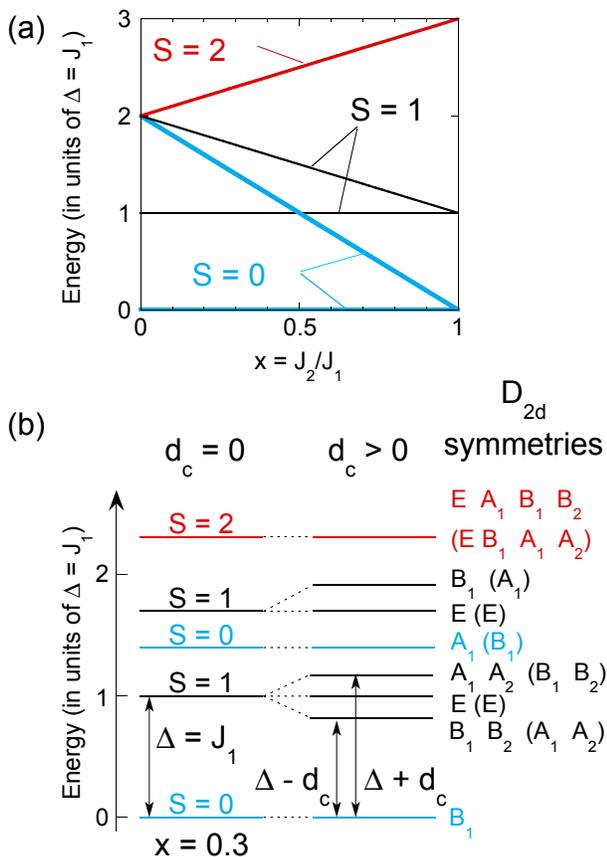,width=80mm}
}
\caption{
(Color online)
(a) The eigenvalues of the H$_{0}$ term of the spin Hamiltonian of 
Eq.~\ref{eq1} corresponding to the 4-spin cluster shown in Fig.~1b.
The spin gap excitation at $\Delta = J_{1}$ is six times degenerate 
in this approximation.
(b) Symmetry analysis of the 4-spin cluster in (a) in the $D_{2d}$ group.
The energies on the left are calculated for a ratio $x = J_{2} / 
J_{1}$ equal to 0.3.
On the right we show the energy splittings and the absolute and 
relative to the ground state (in parenthesis) symmetries of the 16 
magnetic modes when a finite inter-dimer DM interaction $d_{c}$ (see 
Ref.~\cite{CepasPRL01}) is present.
The effect of a finite intra-dimer DM perturbation, $d_{ab}$, is to 
further split the modes which belong to the one-dimensional 
representations.
}
\label{f4}
\end{figure}

If the picture of real space localized elementary triplets is true, 
then one expects that the analysis of the 4-spin cluster in Fig.~1b 
forming the unit cell is able to predict correctly at least some of 
the experimentally observed symmetries of these excitations.
Besides symmetry analysis, numerical diagonalization of the 
Hamiltonian of Eq.~(\ref{eq1}) allows one in principle to also 
identify the energy as well as the predominant spin character of each 
eigenstate of this cluster.
Indeed, as can be seen by comparing Figs.~3 and 4b and looking at the 
rightmost column of the Table~\ref{modes}, the symmetries of the 
observed zero field one-triplet excitations correspond to the results 
of group theory analysis.
The doubly degenerate E modes, T$_{1c}$ and T$_{1d}$, are not 
observed in zero field when the light propagates parallel to the 
$c$-axis because they are accessible only in $(ca)$ or $(cb)$ 
polarizations.
The observation of the T$_{1e}$ (T$_{1f}$) modes with B$_{1}$ 
(B$_{2}$) symmetries at an energy of 2.8~\cm-1 above the A$_{2}$ 
symmetric mode T$_{1b}$ allows the determination of the magnitude of 
the inter-dimer interaction $d_{c}$ (see Fig.~4b) and also of its 
absolute sign \cite{sign}.
The fully symmetric T$_{1a}$ mode which, within the spin model 
including only the $H_{0}$ and $H^{DM}_{c}$ terms given in 
Eqs.~(\ref{eq2}) and (\ref{eq3}), should be degenerate with the 
strong T$_{1b}$ (A$_{2}$ symmetric) excitation at 22.8~\cm-1 is also 
not observed which is most probably due to a much weaker coupling to 
light in this symmetry channel.
We will discuss the coupling mechanisms in the last section of our paper.
\begin{table}[t]
\caption[]
{
Collective spin excitations in zero field: notation, the predominant 
spin character ($S_{tot}$), the $z$ projection of the spin ($S_{z}$), 
the energies and transition symmetries as observed experimentally and 
predicted from the 4-spin cluster in Fig.~1 corresponding to $k = 0$ 
excitations.
T$_{1a}$ ... T$_{1f}$ represent elementary triplet excitations.
Generally, the modes whose energies change in external magnetic 
fields are indexed by~T.
}
\vspace{0.2cm}
\begin{tabular}{c|c|c|ccc}
Mode & S$_{tot}$ (S$_{z}$) & Energy & \multicolumn{3}{c}{Symmetry} \\
   & & & Experiment & & Group Theory \\
\hline
T$_{1a}$ & $1$ ($\pm 1$) & 22.8 & - & \vline & A$_{1}$ \\
T$_{1b}$ & $1$ ($\pm 1$) & 22.8 & A$_{2}$ & \vline & A$_{2}$ \\
T$_{1c}$ & $1$ ($0$) & 24.2 & - & \vline & E \\
T$_{1d}$ & $1$ ($0$) & 24.2 & - & \vline & E \\
T$_{1e}$ & $1$ ($\pm 1$) & 25.6 & B$_{1}$ & \vline & B$_{1}$ \\
T$_{1f}$ & $1$ ($\pm 1$) & 25.6 & B$_{2}$ & \vline & B$_{2}$ \\
\hline
S$_{1}$ & 0 ($0$) & 28.9 & B$_{1}$ & \vline & - \\
T$_{2}$ & 1 & 37.5 & A$_{2}$ & \vline & - \\
T$_{3}$ & 1 & 40.8 & A$_{2}$ & \vline & - \\
S$_{2}$ & 0 & 43.0 & B$_{1}$ & \vline & - \\
T$_{4}$ & 1 & 44.5 & A$_{2}$ & \vline & - \\
S$_{3}$ & 0 & 45.2 & B$_{1}$ & \vline & - \\
T$_{5}$ & 1 & 50.9 & A$_{2}$ & \vline & - \\
\end{tabular}
\label{modes}
\end{table}

We turn now to the discussion of two-triplet states.
Besides the singlet ground state and six one-triplet states, as can 
be seen in Fig.~4, the 4-spin cluster analysis predicts the following:
one $S = 0$ two-triplet bound state in the B$_{1}$ channel, three 
branches with A$_{1}$ and doubly degenerate E symmetries which belong 
to another bound $S = 1$ excitation and five branches of a quintuplet 
($S = 2$) state having A$_{1}$, A$_{2}$, B$_{1}$ and E symmetries 
with respect to the ground state.
Their energies are plotted in Fig.~4a as a function of $x = J_{2}/J_{1}$.
We note that due to symmetry reasons none of the observed A$_{2}$ symmetric 
modes from T$_{2}$ to T$_{5}$, having energies higher than 30~\cm-1, 
qualify for an interpretation as triplet bound states generated 
within the 4-spin cluster.
This is consistent with the fact that larger cluster sizes are 
necessary in order to capture the more delocalized nature of these 
excitations.

The fact that the existence of the strong A$_{2}$ symmetric bound 
triplet state at an energy 1.55~$\cdot \Delta$~=~37.5~\cm-1 has not 
been predicted by high order perturbative analysis~\cite{KnetterPRL00} 
(here we refer especially to the symmetry of this excitation, not its 
energy) suggests that other spin interactions have to be taken into 
account in order to explain the excitation spectrum.
Apparently the symmetry considerations would allow the 28.9~\cm-1 
feature denoted by S$_{1}$ in Fig.~3 to be interpreted as the singlet 
bound state of two triplets within a unit cell.
As we show in the following section, the 28.9~\cm-1 mode does not 
shift in external fields, which is compatible with a collective 
singlet excitation as discussed in~\cite{LemmensPRL00}, but suggests 
that its internal structure is not the one derived from the 4-spin 
cluster.

\section{Magnetic field effects at T~$\approx$~3~K}
In Fig.~5, using the same mode notations, we show the influence of an 
external magnetic field applied parallel and perpendicular to the 
$c$-axis on the low temperature Raman spectra from Fig.~3.
Here we summarize the relevant aspects, noting first that energy 
shifts induced by magnetic fields in this figure were observed only 
for the modes indexed by T in the Table~\ref{modes}.
In Fig.~5a we observe the splitting of the T$_{1a}$ and T$_{1b}$ 
modes in magnetic fields $\vec{B} \parallel \hat{c}$, the B~=~1~T 
spectrum showing that the A$_{2}$ mode (T$_{1b}$) present in zero 
field disperses upwards with increasing magnetic field.
Dashed lines mark the dispersion of the much weaker modes T$_{3}$, 
T$_{4}$ and T$_{5}$.
In Fig.~5b one of the E modes at 24.2~\cm-1 becomes Raman active due 
to symmetry lowering for $\vec{B} \perp \hat{c}$ configuration and we 
observe three dispersing branches of the gap multiplet.
Fig.~5c shows that the B$_{1}$ symmetric excitation at 28.9~\cm-1 
does not change its energy with field, only a very small negative 
shift of the order of 0.5~\cm-1 from 0 to 6~T is seen because of the 
crossing with the upward dispersing gap branches seen in $(RR)$ 
polarization.
Fig.~5d, which is a zoomed in region of Fig.~5b, shows that several 
modes become Raman active in finite fields $\vec{B} \perp \hat{c}$ 
around 38~\cm-1 where the  T$_{2}$ excitation lies.
The internal structure of this higher energy multiplet is composed of 
modes dispersing up, down and independent of magnetic field.
We remark on the similarity in selection rules and dynamics in 
magnetic fields between the collective modes around 38~\cm-1 and that 
of the spin gap branches around 24~\cm-1.
The emergence in finite fields of several strong modes in the spin 
gap region precludes the observation of the dynamics of the weak 
T$_{1e}$ and T$_{1f}$ modes from Fig.~3.

Fig.~6 summarizes the magnetic field dependencies of the energies and 
spectral weights of the most intense Raman excitations.
The symbols in Fig.~6a-d correspond to the experimental data and the 
solid lines are results of calculations: the energies in panels a and 
b are obtained by exact diagonalization of the Hamiltonian of 
Eq.~\ref{eq1} with the parameters specified in the next paragraph; 
using Fermi's golden rule, the intensities in Fig.~6c and Fig.~6d are 
calculated as the square of the matrix elements between the ground 
and excited states of the effective Fleury-Loudon spin interactions 
describing the coupling to the external electromagnetic 
field~\cite{FleuryPR68}.
The form of these interaction terms, denoted by $H^{int}_{1}$ and 
$H^{int}_{2}$, are discussed explicitly in the next section.
\begin{figure}[t]
\centerline{
\epsfig{figure=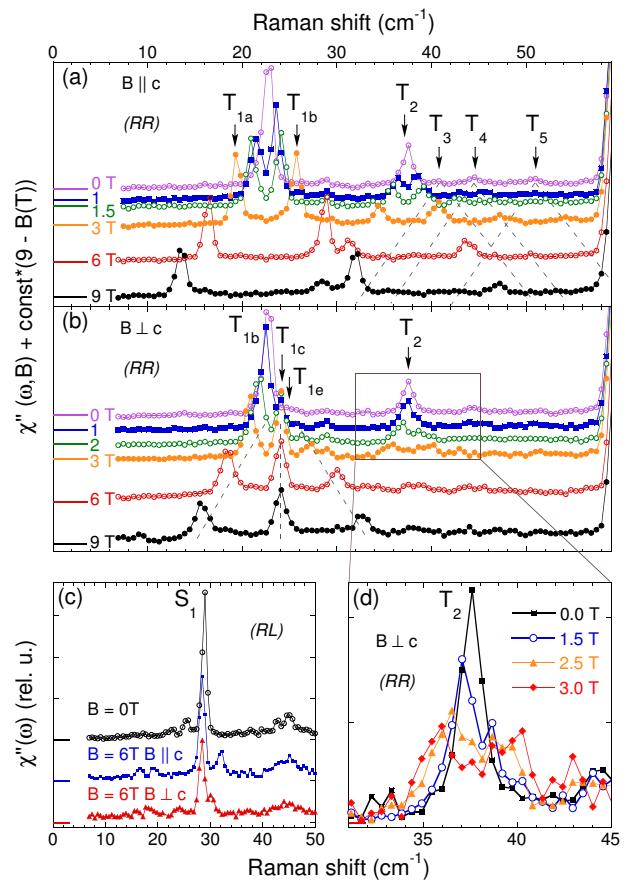,width=80mm}
}
\caption{
(Color online)
Magnetic field dependencies of the magnetic excitations at T~=~3~K 
using the $\omega_{L}$~=~1.92~eV excitation energy in the following 
geometries:
(a) $(RR)$ $\vec{B} \parallel \hat{c}$.
(b) $(RR)$ $\vec{B} \perp \hat{c}$.
In (c) the $(RL)$ polarized data is shown for 0 and 6~T magnetic 
fields for both $\vec{B} \parallel \hat{c}$ and $\vec{B} \perp 
\hat{c}$.
In (a) and (b) the vertical shift is proportional to the magnetic 
field difference with respect to the 9~T spectrum and the dashed 
lines are guides for the eye.
Panel (d) is a zoomed in region of panel (b).
}
\label{f5}
\end{figure}

Here we discuss the parameter set used for data fitting in both the 
$\vec{B} \parallel \hat{c}$ and $\vec{B} \perp \hat{c}$ 
configurations.
Taking into account that: (i) the 4-spin cluster neglects many-body 
gap renormalization effects leading to a singlet-triplet energy 
independent of $x = J_{2}/J_{1}$, see Fig.~4a, and (ii) the fact that 
when using periodic boundary conditions there is an effective 
doubling of the $J_{2}$ and inter-dimer DM interactions, we chose the 
following values:
$J_{1}$~=~$\Delta$~=~24.2~\cm-1 which is the value of the spin gap 
(see the Table and 
Refs.~\cite{KageyamaPRL00,CepasPRL01,NojiriJPSJ99,NojiriJPSJ03,ToomasPRB00}); 

$x = 0.556$ from the ratio of the energies of the sub-gap mode at 
21.5~\cm-1, Refs.\cite{NojiriJPSJ03,ToomasPrivate}, with respect to 
the gap $\Delta$ (see Fig.~4a);
an inter-dimer DM term $d_{c}$~=~1.4~\cm-1, which produces the 
splitting of the T$_{1a,b}$ and T$_{1e,f}$ branches from 24.2~\cm-1 
to 22.8 and 25.5~\cm-1 respectively (our value of $d_{c}$ is 
consistent with the one proposed in Ref.~\cite{CepasPRL01});
finally, using the magnetic field value where the intensities cross 
in Fig.~6c, the value of the intra-dimer interaction was set to 
$d_{ab}$~=~2.66~\cm-1.
We note two aspects regarding the magnitude of the parameters used above.
The first is that because many-body spin interactions are not 
captured within this minimal 4-spin cluster, our chosen value for $x 
= J_{2}/J_{1}$ should not be taken \emph{ad litteram} for the real 
structure.
The second aspect, discussed in more detail later in this section in 
connection to the dispersion of the magnetic modes in the $\vec{B} 
\parallel \hat{c}$ configuration, is related to the fact that we were 
forced to choose $x > 0.5$, which is equivalent to saying that the 
singlet bound state has an energy which is lower than the elementary 
gap excitation at 24.2~\cm-1.
This fact may be important regarding the experimental observation of 
a mode at 21.5~\cm-1, see 
Refs.~\cite{CepasPRL01,NojiriJPSJ03,ToomasPrivate}.
\begin{figure}[t]
\centerline{
\epsfig{figure=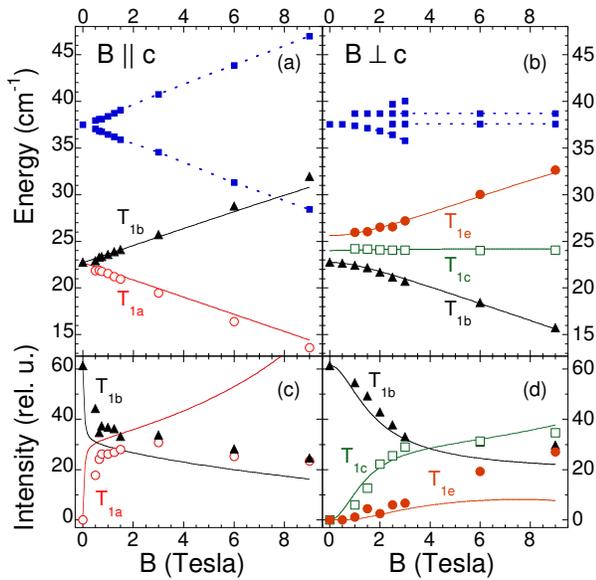,width=80mm}
}
\caption{
(Color online)
Energies (panels a and b) and intensities (panels c and d) of the 
spin excitations for $\vec{B} \parallel \hat{c}$ (left) and $\vec{B} 
\perp \hat{c}$ (right) from Fig.~5a-b.
Symbols represent experimental points, solid lines are the results of 
4-spin cluster diagonalization as described in the text, dashed lines 
are guides for the eye.
}
\label{f6}
\end{figure}

The space group symmetry of the crystal uniquely imposes for any 
existing static intra-dimer DM interaction (the term $H^{DM}_{ab}$ in 
Eq.~\ref{eq4}) the configuration depicted in Fig.~1a, i.e. the DM 
vectors are perpendicular to the $c$-axis and dimer bonds.
It is the $H^{DM}_{ab}$ term in the system Hamiltonian which is 
responsible in our interpretation for the mixing of singlet and 
triplet modes and allows for a finite coupling of the latter 
excitations to the external photon field.
As we mentioned in the introduction, a finite intra-dimer interaction 
$d_{ab}$ is crucial because the $H^{DM}_{c}$ term alone does not mix 
the singlet and triplet states.
Also shown in Fig.~6 by filled squares and dashed lines are the 
experimental field dependencies of other higher energy modes observed 
in Fig.~4.

The agreement for this set of parameters is qualitatively good 
overall and quantitatively better with regard to the energy and 
intensity variations for the $\vec{B} \perp \hat{c}$ case.
As discussed before, the term $H^{DM}_{ab}$ plays a crucial role in 
obtaining a finite coupling to the excited $S = 1$ triplets.
$H^{DM}_{ab}$ also produces splittings of the T$_{1a}$ and T$_{1b}$ 
modes, of the T$_{1e}$ and T$_{1f}$ modes as well as that of the 
quintuplet branch (these splittings are not shown explicitly in 
Fig.~4b).
For the spin gap branches, the magnitude of the splittings is 
unresolved because it is very small, of the order of $d_{ab}^{\ 2} / 
\Delta \approx$~0.25~\cm-1.
The largest discrepancy between the experimental data and the 
calculation is seen in Fig.~6c.
One aspect in this regard is that the value $d_{ab}$ had to be chosen 
greater than that of $d_{c}$.
This is unexpected because $d_{c}$ is allowed by symmetry both above 
and below the structural phase transition at 395~K while the 
existence of a finite intra-dimer DM interaction is allowed only 
below 395~K, when the mirror symmetry of the $(ab)$ plane is broken.
Additional terms may be responsible for this disagreement, possible 
candidates being the in-plane components of the inter-dimer DM 
interaction, which should also be allowed below the structural phase 
transition.

We now discuss the existence of a magnetic mode \emph{below} the spin 
gap value~\cite{CepasPRL01,NojiriJPSJ03,ToomasPrivate}.
In order to reproduce the upward dispersion with fields $\vec{B} 
\parallel \hat{c}$ of the T$_{1b}$ mode we had to choose a value for 
$x$ which is greater than 0.5, otherwise this excitation would have 
displayed a downward dispersion.
>From Fig.~4a we observe that $x = J_{2}/J_{1} \geq$~0.5 implies that 
>the position of the bound singlet state is below $\Delta$.
We suggest that precisely this state may be responsible for the 
observations of the 21.5~\cm-1 mode in 
Refs.~\cite{CepasPRL01,NojiriJPSJ03,ToomasPrivate}.
The presence of this excitation will also influence specific heat 
measurements and, in conjunction with the finite intra-dimer 
interaction $d_{ab}$, also the low temperature magnetization data 
which is not quantitatively understood yet~\cite{MiyaharaJPCM03}.

The existence of this magnetic mode below the gap seems 
quantitatively at odds with theoretical 
predictions~\cite{KnetterPRL00,MiyaharaJPSJS00,MiyaharaPRL99}.
Nevertheless, perturbational calculations predict the existence of a 
singlet state at 25~\cm-1, which is above but very close to the spin 
gap~\cite{KnetterPRL00}.
A quantitative reconciliation between theory and the observed 
selection rules in magnetic fields could be achieved if the coupling 
ratio $x~=~J_{2}/J_{1}$ is slightly increased from the value of 
0.603, as determined in Ref.~\cite{KnetterPRL00}.
The experimental finding of the set of A$_{2}$ modes (T$_{2}$ to 
T$_{5}$ in Fig.~3, all of them below the two-magnon continuum 
starting at 2$\Delta \approx$~48~\cm-1 and whose symmetries are also 
not predicted by theory) shows that although several aspects of the 
magnetic bound states are understood, a complete picture of the spin 
dynamics in the multi-triplet sectors is still to be achieved.

\section{Resonance and light coupling mechanisms}
Fig.~7 shows two low temperature Raman spectra taken in $(a'b')$ 
polarization with two incoming laser frequencies, $\omega_{L}$~=~1.92 
and 2.6~eV.
The point we want to make in this paragraph is that we observe two 
qualitatively different behaviors.
Firstly we notice that the area under the peak corresponding to the 
T$_{1b}$ mode in the spectrum taken with the excitation energy 
$\omega_{L}$~=~1.92 is of the same order of magnitude as the one in 
the spectrum taken with the 2.6~eV laser excitation frequency.
The same observation is also true for the mode denoted by T$_{2}$.
>From this perspective, a different behavior is observed for other 
>modes and an example is the group of modes denoted by T$_{1e}$, 
>S$_{1}$, S$_{2}$ and S$_{3}$.
In this latter case for instance, it is clearly seen in our spectra 
that the mode S$_{1}$ in the spectrum taken with the 
$\omega_{L}$~=~2.6~eV excitation is several orders of magnitude 
stronger than when using the $\omega_{L}$~=~1.92~eV laser line.
Also, while the modes T$_{1e}$, S$_{2}$ and S$_{3}$ are barely 
visible when $\omega_{L}$~=~1.92~eV (see also Fig.~3), they become 
quite strong for $\omega_{L}$~=~2.6~eV.
Fig.~5 shows in addition that other modes in the 50 to 70~\cm-1 
energy range also become visible only in the $\omega_{L}$~=~2.6~eV 
spectrum.
We conclude this paragraph by saying that the group 
\{T$_{1b}$,T$_{2}$\} has a different resonance behavior than the 
group \{T$_{1e}$,S$_{1}$,S$_{2}$,S$_{3}$\} in the sense that their 
relative intensities are quite different for the two laser excitation 
frequencies used.
\begin{figure}[t]
\centerline{
\epsfig{figure=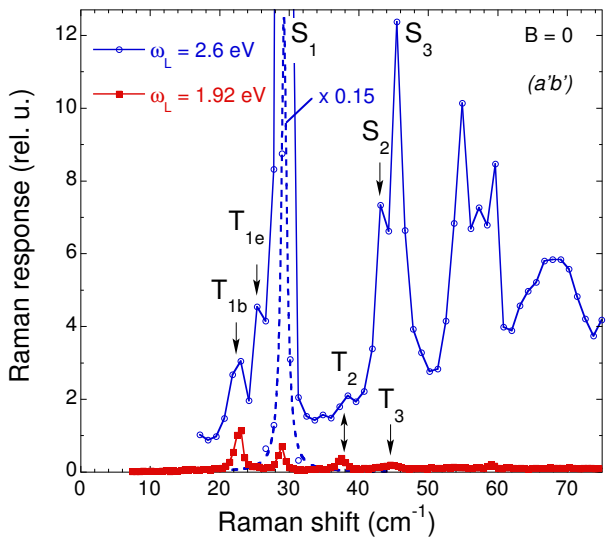,width=80mm}
}
\caption{
(Color online)
T~=~3~K Raman data in $(a'b')$ polarization for $\omega_{L}$~=~1.92 
(filled squares) and 2.6~eV (empty circles) excitation energies.
The data points corresponding to the resonantly enhanced mode at 
29~\cm-1 are multiplied by 0.15 and the corresponding line represents 
a Lorentzian fit.
}
\label{f7}
\end{figure}

Note that in the paragraph above we commented only on the 
\emph{relative} intensities of the Raman modes when the 
$\omega_{L}$~=~1.92~eV excitation was used compared to the 2.6~eV 
line.
Although the absolute intensities of each the Raman excitations can 
be affected by optical corrections of the data, this effect cancels 
out when relative intensity ratios are considered.
In other words, it is possible in principle that optical corrections 
can change the ratio of the absolute values of the Raman intensities 
of the T$_{1b}$ or T$_{2}$ modes when taken with $\omega_{L}$~=~1.92 
and 2.6~eV laser energies respectively.
The same observation applies when we relatively compare the 
intensities of the T$_{1e}$, S$_{1}$, S$_{2}$ and S$_{3}$ for these 
two excitations.
However, the important point is that the ratio of intensities should 
be affected in exactly the same way \emph{if the same resonance 
mechanism was responsible for both the \{T$_{1b}$,T$_{2}$\} group of 
excitations on one hand and for the group of modes 
\{T$_{1e}$,S$_{1}$,S$_{2}$,S$_{3}$\} on the other hand}.
The conclusion following from the above discussion is that the data 
in Fig.~7 proves the existence of two light coupling mechanisms to 
the observed low energy excitations.
In fact, further support for our statement comes from preliminary 
data which show that the 1.92 and 2.6~eV laser excitations energies 
correspond to regions in the visible range where distinct features of 
the reflectivity spectrum are observed~\cite{Sasa}.
While certainly interesting in this respect, a complete determination 
of the optical parameters of \scbo in the visible range and the 
identification of the specific high energy electronic intermediate 
states involved in the coupling mechanisms goes beyond the scope of 
our article and, as explained above, cannot change our conclusions.

The resonance of the B$_{1}$ symmetric T$_{e}$ magnetic mode, 
enhanced for $\omega_{L}$~=~2.6~eV, is similar to the one 
corresponding to S$_{1}$, S$_{2}$ and S$_{3}$ excitations as well as 
to the behavior of the new modes seen around 55, 59 and 68~\cm-1.
The results of perturbational analysis regarding energy scales and 
symmetries~\cite{KnetterPRL00}, in particular of a B$_{1}$ symmetric 
$S~=~0$ states at 45~\cm-1, argues for the magnetic nature of the 
mode S$_{3}$,~Ref.\cite{LemmensPRL00}.
Just on the account of the similar resonance behavior of the S$_{3}$ 
excitation with respect to the modes found above 50~\cm-1 one cannot 
unambiguously identify the latter as magnetic bound states as well.
The lack of energy shift in magnetic fields is consistent with such 
an interpretation, but the same would be true if they had a phononic 
origin.

We discuss below the nature of the two light coupling mechanisms to 
magnetic excitations.
For the set of modes discussed in Fig.~6c-d we propose that the 
coupling takes place $via$ the spin-orbit interaction which can be 
written in an effective form as $H^{int}_{1} \propto ({\bf e}_{in} 
\times {\bf e}_{out}) {S}_{tot}^{\ z}$ (Ref.~\cite{FleuryPR68}).
As it is found experimentally, in zero field this interaction 
Hamiltonian probes indeed excitations with A$_{2}$ symmetry (the 
T$_{1b}$ mode) and the calculated magnetic field dependent 
intensities for this and other modes in Fig.~6c-d is also in 
agreement with the experimental results.
The coupling to the T$_{1e}$ and T$_{1f}$ modes from Fig.~3 can be 
possibly understood if we invoke the usual effective spin interaction 
corresponding to the photon induced spin exchange process 
$H^{int}_{2} \propto \sum_{<i,j>} ({\bf e}_{in} {\bf r}_{ij}) ({\bf 
e}_{out} {\bf r}_{ij}) {\bf S}_{i} \cdot {\bf S}_{j}$.
Here the sum runs over pairs of lattice sites, ${\bf S}_{i}$ and 
${\bf S}_{j}$ are the exchanged spins on sites $i$ and $j$ 
respectively, while {\bf r}$_{ij}$ is the vector connecting these 
sites~\cite{FleuryPR68}.
The explicit expression of this interaction for several polarizations 
in the 4-spin cluster approximation contains finite coupling in 
B$_{1}$ and B$_{2}$ channels for the triplet T$_{1e}$ and T$_{1f}$ 
states.
This explains the presence of the 25.6~\cm-1 magnetic modes in all 
polarizations except~$(RR)$.

The difference in the coupling strengths seen in Fig.~7 is thus 
understandable because these two light coupling mechanisms given by 
$H^{int}_{1}$ and $H^{int}_{2}$ need not be simultaneously in 
resonance with the same high energy excited electronic states.
The photon induced spin exchange Hamiltonian, $H^{int}_{2}$, is 
usually invoked in order to explain Raman active $S = 0$ two-magnon 
type excitations in various magnetic systems~\cite{FleuryPR68}.
\scbo is an example where this Hamiltonian, in the presence of 
singlet-triplet mixing DM interactions, can be used to account for 
coupling to $S = 1$ states.
We also note that in principle the photon induced spin exchange could 
provide coupling to the 21.5~\cm-1 $S = 0$ bound state below the 
spin gap.
The reason this mode is not directly observed in our spectra for any 
of the two excitation energies used in Fig.~6 is an open question.
However, one possible explanation is that the Raman form factor for 
exciting pairs of $k = 0$ magnons is vanishing as opposed to the case 
zone boundary modes.
An example is the case of the Raman vertex calculated for the 2D 
square lattice within the spin-wave approximation~\cite{SandvikPRB98} 
and using the Fleury-Loudon~\cite{FleuryPR68} interaction.
Consequently, both the 21.5 and the 28.9~\cm-1 excitations could be 
attributed to $S = 0$ bound states but originating from different 
parts of the Brillouin zone and having substantially different 
binding energies.

\section{Summary}
We study by Raman scattering collective magnetic excitations in the 
spin-dimer compound \scbo.
Regarding the one-triplet sector, we showed that by using a 4-spin 
cluster approximation and by including an additional intra-dimer DM 
interaction we are able to explain the observed zero field symmetry 
selection rules and the rich behavior in magnetic fields.
We are also able to experimentally demonstrate the existence of two 
effective magnetic light scattering Hamiltonians which are 
responsible for their resonance behavior.
The 4-spin approximation fails to account for the excitations seen in 
the multi-particle magnetic sectors.
In particular, the existence of a set of four modes below the onset 
of the two-triplet continuum (at 37.5, 40.8, 44.5 and 50.9~\cm-1 in 
the A$_{2}$ symmetry channel) shows that further theoretical analysis 
is required in order to understand the nature of these composite 
excitations.
We suggest a possible explanation for the existence of a sub-gap 
collective mode in terms of a strongly bound singlet state which can 
be generated within the space of 4 nearest neighbor spin dimers.

Experimental data in the energy range below 350~\cm-1 shows the 
existence of several quasi-degenerate phonons.
General symmetry arguments suggest that these excitations, which are 
very close in energy, involve different vibrational patterns (in 
plane and $c$-axis atomic motions respectively).
The failure of group theory to provide an understanding of this 
interesting behavior at a qualitative level calls for a detailed 
theoretical investigation of the vibrational modes.
We also find a weak A$_{2}$ symmetric phonon at 155.9~\cm-1.
The energy and symmetry of this excitation suggest that this is a 
coupled spin-phonon excitation and can be evidence for the existence 
of magneto-elastic interactions in \scbo.

We thank T. R\~o\~om, G.S. Uhrig, T. Siegrist and R. Stern for 
discussions. A.G. also gratefully acknowledges the collaboration with Sasa 
Dordevic.



\begin{references}
\bibitem{sign}
A sign change will interchange the position of these modes around the 
gap $\Delta$.
\end{references}
\end{document}